\documentclass[aps,prl,twocolumn,preprintnumbers,amsmath,amssymb,superscriptaddress,10pt]{revtex4-2}
\usepackage{amsmath,graphicx}
\usepackage[utf8]{inputenc}
\usepackage[T1]{fontenc}
\usepackage{xcolor}
\usepackage{textcomp}
\usepackage{bm}
\usepackage{physics}
\usepackage{amsmath}
\usepackage{tikz}
\usepackage{mathdots}
\usepackage{yhmath}
\usepackage{cancel}
\usepackage{color}
\usepackage{array}
\usepackage{multirow}
\usepackage{amssymb}
\usepackage{gensymb}
\usepackage{tabularx}
\usepackage{extarrows}
\usepackage{booktabs}
\usetikzlibrary{fadings}
\usetikzlibrary{patterns}
\usetikzlibrary{shadows.blur}
\usetikzlibrary{shapes}

\usepackage{color}
\definecolor{LinkColor}{rgb}{0.75,0.0,0.2}

\usepackage{hyperref}
\hypersetup{
	pdfauthor={good guys},
	pdftitle={good title},
	colorlinks=true,
	citecolor=LinkColor,
	linkcolor=LinkColor,
	urlcolor=LinkColor,
}

\usepackage{listings}
\definecolor{lightgray}{gray}{1}

\usepackage{theorem}

\usepackage{tikz}

\newcommand{\nc}{\newcommand}
\nc{\braoprket}[3]{\langle#1|#2|#3\rangle}
\nc{\opn}[1]{\operatorname{#1}}
\nc{\avg}[1]{\langle#1\rangle}
\nc{\ketbrasame}[1]{|#1\rangle\!\langle#1|}
\nc{\swap}{\opn{SWAP}}
\nc{\E}{\mathbb{E}}
\nc{\Var}{\opn{Var}}
\nc{\dg}{\dagger}

\usepackage[normalem]{ulem}

\begin{document}

\title{Symmetry-Enriched Criticality from Kennedy–Tasaki Self-Duality} 

\author{Mingyuan Chen}
\altaffiliation{The first two authors contributed equally.}
\affiliation{Eastern Institute of Technology, Ningbo 315200, China}

\author{Sheng Yang}
\altaffiliation{The first two authors contributed equally.}
\affiliation{Institute for Advanced Study in Physics and School of Physics, Zhejiang University, Hangzhou 310058, China}

\author{Xue-Jia Yu}
\email{xuejiayu@eitech.edu.cn}
\affiliation{Eastern Institute of Technology, Ningbo 315200, China}

\begin{abstract}

The interplay between topology and quantum criticality has established symmetry-enriched criticality as a new paradigm beyond conventional frameworks in condensed matter and statistical physics. However, the conditions under which such criticality emerges remain poorly understood.
In this work, we show that a broad class of symmetry-enriched critical points separating symmetry-protected topological (SPT) and spontaneous symmetry-breaking (SSB) phases is governed by Kennedy--Tasaki (KT) self-duality.
At these criticalities, the KT transformation that interchanges the SPT and SSB phases acts as a self-dual noninvertible symmetry, endowing the critical points with nontrivial topological characteristics.
We numerically demonstrate this mechanism in $\mathbb{Z}_{n} \times \mathbb{Z}_{n}$ cluster-qudit chains and further argue that it naturally extends to higher dimensions, providing a duality-based criterion for identifying symmetry-enriched criticality across dimensions.

\end{abstract}

\maketitle

\emph{\color{red} Introduction.}---Understanding the mechanisms underlying the emergence of quantum critical points (QCPs) is a central theme in modern condensed matter and statistical physics~\cite{landau2013statistical,Sachdev_1999,cardy1996scaling,Sondhi1997RMP}.
Among the various theoretical frameworks, duality has long served as a powerful organizing principle for understanding QCPs~\cite{SENTHIL20191,SEIBERG2016395,KW1941PR1,KW1941PR2,Shimshoni1997PRB,Wang2017PRX,Qin2017PRX,Somoza2021PRX,Lu2021PRB,Li2022PRB,Yang2023PRB,Yang2023PRB_b,Lootens2023PRXQuantum}.
A paradigmatic example is the Kramers--Wannier (KW) duality~\cite{KW1941PR1,KW1941PR2} of the $(1+1)$-dimensional transverse-field Ising model, which exchanges the Ising-interaction and transverse-field terms and maps the spontaneous symmetry-breaking (SSB) and symmetry-preserving phases onto each other.
Consequently, the parameter point that is invariant under the KW duality must be critical.
This classic result has recently gained a deeper interpretation within the framework of noninvertible symmetry~\cite{shao2024whatsundonetasilectures,SCHAFERNAMEKI20241,Choi2024JHEP,Chatterjee2024SciPost,Warman2025PRL,Bhardwaj2025Scipost,Zhang2025JHEP,Chung2026PRL,Li2026PRL,Bhardwaj2026SciPost,Su2026SciPost,chen2026spontaneousbreakingnoninvertiblesymmetries}: 
the KW duality can be viewed as a noninvertible symmetry with a quantum anomaly, which pins the system to the boundary between the two duality-related phases and thereby enforces a gapless critical point~\cite{Chatterjee2024SciPost,Bhardwaj2026SciPost,Li2026PRL}.

For decades, universality classes of critical points were believed to be uniquely characterized by critical exponents, providing the fundamental framework for classifying critical phenomena~\cite{landau2013statistical,Sachdev_1999,cardy1996scaling,Sondhi1997RMP}.
This conventional paradigm has recently been challenged by the discovery of topological physics in quantum critical systems~\cite{YU20261}.
Specifically, symmetry-protected topological (SPT) edge modes—previously believed to exist only in gapped systems—have been shown to coexist with a gapless bulk at criticality, leading to topologically distinct universality classes even among systems that share the same critical exponents, a phenomenon now referred to as symmetry-enriched quantum criticality~\cite{Verresen2018PRL,verresen2020topologyedgestatessurvive,Verresen2021PRX,Duque2021PRB,Yu2022PRL,Mondal2023PRB,Yu2024PRB,Prembabu2024PRB,Zhong2025PRB,Huang2025SciPost,Li2025PRB,Rey2025PRB,Zhou2025CP,chou2025ptsymmetryenrichednonunitarycriticality,guo2026generalizedlihaldanecorrespondencecritical,deng2026anomalousdynamicalscalingtopological,prembabu2025noninvertibleinterfacessymmetryenrichedcritical,yang2026topologicalquantumcriticalityquasiperiodic,TanCP2026,yang2026topologicaltricriticalisinguniversality,chou2026topologicallyenforcedlifshitzmulticriticality,Cheng2026Nature,Lin2026Nature}.
This conceptual extension not only generalizes the notion of topology to gapless quantum critical systems, but also reveals fundamentally novel phenomena beyond conventional critical theory and topological physics~\cite{Keselman2015PRB,Scaffidi2017PRX,JIANG2018753,Parker2018PRB,Thorngren2021PRB,Hidaka2022PRB,Wen2023PRB,Yu2024PRL,Zhang2024PRA,Su2024PRB,Li2025SciPost,Wen2025PRB,Yang2025CP,Flores2025PRL,yang2025deconfinedcriticalityintrinsicallygapless,prembabu2025multicriticalitypurelygaplessspt,Yu2026PRB,xu2026frameworkpredictingentanglementspectra,chou2026topologicallyenforcedlifshitzmulticriticality}.
Despite this progress, the criteria governing the emergence of symmetry-enriched quantum criticality remain to be established and constitute an outstanding question.

\begin{figure}[t]
  \includegraphics[width=\linewidth]{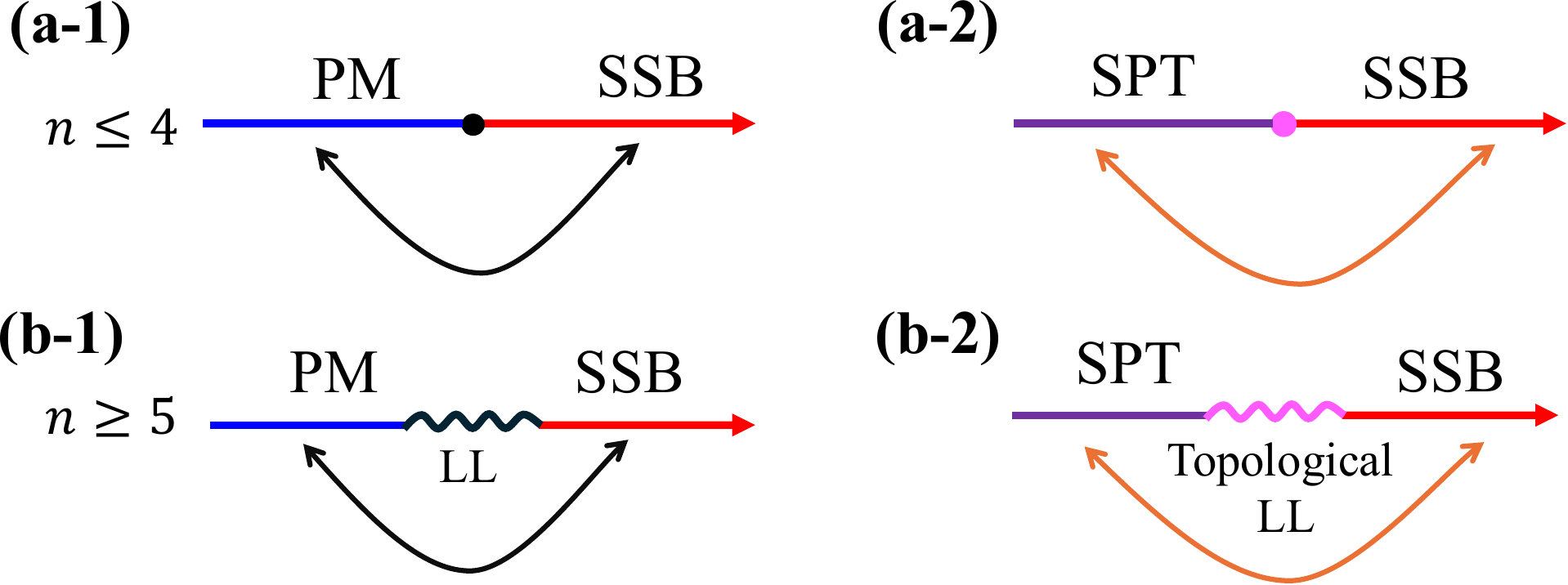}
  \caption{Schematic phase diagrams of the $\mathbb{Z}_{n} \times \mathbb{Z}_{n}$ qudit chains. 
  The left panels [(a-1) and (b-1)] illustrate the quantum phase transitions between the paramagnetic (PM) and spontaneous symmetry-breaking (SSB) phases.
  These two phases are related by the Kramers--Wannier (KW) duality, denoted by the black solid arrows. The right panels [(a-2) and (b-2)] depict the transitions between the symmetry-protected topological (SPT) and SSB phases, which are connected via Kennedy--Tasaki (KT) duality, indicated by the orange solid arrows.
  For $2 \leq n \leq 4$ [(a-1) and (a-2)], the system undergoes a direct continuous topologically trivial (nontrivial) phase transition at a single critical point, while for $n \geq 5$ [(b-1) and (b-2)], the transition broadens into an intermediate extended gapless region, realizing a standard (topological) Luttinger liquid (LL).}
  \label{fig:schematic}
\end{figure}

In this work, we fill this gap by uncovering a duality-based criterion  that triggers symmetry-enriched criticality between SPT and SSB phases.
Building on the modern understanding of the KW self-duality as a noninvertible symmetry that enforces conventional criticality, we demonstrate that the Kennedy--Tasaki (KT) self-duality likewise defines a noninvertible symmetry, but one that additionally endows the critical point with nontrivial topological properties, thereby giving rise to symmetry-enriched criticality.
We substantiate this mechanism through comprehensive lattice simulations of $\mathbb{Z}_{n} \times \mathbb{Z}_{n}$ cluster-qudit chains and further argue that the underlying principle naturally extends to higher dimensions, where reliable numerical simulations remain highly challenging.
These results establish KT self-duality as a universal organizing principle for investigating symmetry-enriched criticality.

\emph{\color{red} KW self-duality and gapless quantum criticality.}---We begin with the paradigmatic $(1+1)$-dimensional transverse-field Ising chain under periodic boundary conditions (PBC), $H_{\mathrm{TFI}} = -\sum_{i} \left( J_2 Z_{i} Z_{i+1} + h_2 X_{i}\right)$ where $X_i$ and $Z_i$ are Pauli operators acting on site $i$.
The model exhibits two distinct phases: a ferromagnetic (FM) phase with spontaneous $\mathbb Z_2$ symmetry breaking and a symmetric paramagnetic (PM) phase.
These two phases are exchanged by the celebrated KW transformation, which interchanges the Ising interaction and the transverse-field term, $Z_iZ_{i+1}\leftrightarrow X_i$, thereby singling out the self-dual point, $J_2=h_2$, as the unique transition between them.
While the KW transformation is traditionally viewed as a lattice duality relating different descriptions of the model, a modern perspective reveals a deeper structure: precisely at the self-dual point, it is promoted to a non-invertible symmetry.
Unlike an ordinary global symmetry, this transformation combines a unitary transformation with a projection onto the $\mathbb Z_2$-symmetric sector and therefore has no inverse~\cite{Lu2024SciPost}.
The resulting noninvertible symmetry carries an anomaly that forbids a trivially gapped phase, thereby enforcing quantum criticality.

The same structure extends naturally to the $\mathbb Z_n$ clock qudit chain under PBC~\cite{Duivenvoorden2013PRB,Bondesan_2013,Else2013PRB,geraedts2014exactmodelssymmetryprotectedtopological,Li2015PRB,TSUI2017470,Xu2017PRB,Whitsitt2018PRB,Alavirad2021PRB,Karch2025JHEP}, $H_{\mathbb{Z}_n\text{-clock}} = -\sum_{i} \left(J_{n} R^{\dagger}_{i}R_{i+1} + h_{n} M_{i} + \mathrm{h.c.}\right)$, where $R_i$ and $M_i$ denote the generalized Pauli (clock and shift) operators (see End Matter for definitions).
As in the Ising chain, the $\mathbb{Z}_n$ generalized KW transformation~\cite{Su2026SciPost,Li2026PRL,Parayil2024PRB,Lu2024SciPost} exchanges the FM interaction and transverse-field terms, rendering the Hamiltonian self-dual at $J_n=h_n$.
At this point, the $\mathbb{Z}_n$ KW transformation [see Supplemental Material (SM) Sec.~\ref{sm1} for its explicit form] is promoted to a noninvertible symmetry that pins the critical point. 
Remarkably, the way in which this gaplessness is realized depends sensitively on $n$. For $2 \leq n \leq 4$, the self-duality stabilizes an isolated QCP described by the $\frac{SU(2)_n}{U(1)}$ coset conformal field theory (CFT) with central charge $c=\frac{2(n-1)}{n+2}$~\cite{di1997conformal,ginsparg1988appliedconformalfieldtheory,TSUI2017470,Whitsitt2018PRB} [Fig.~\ref{fig:schematic}(a-1)].
By contrast, for $n\ge5$, the same self-duality stabilizes an extended gapless Luttinger-liquid phase with central charge $c=1$ [Fig.~\ref{fig:schematic}(b-1)].
These examples identify the KW self-duality as an organizing principle for conventional quantum criticality and naturally motivate the question of whether an analogous self-duality can likewise enforce symmetry-enriched quantum criticality.

\begin{figure*}[htbp]
  \centering
  \includegraphics[width=0.8\linewidth]{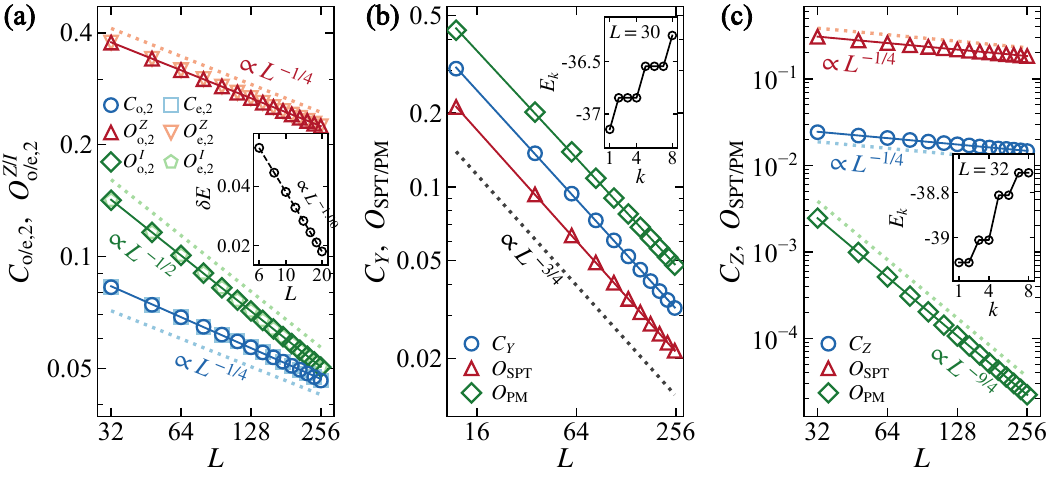}
  \caption{Finite-size scaling of connected spin correlations and PM and SPT string order parameters at three critical spin-$1/2$ chains under open boundary conditions: 
  (a) $H_{2} = -\sum_i \left( Z_{i-1} X_i Z_{i+1} + Z_{i-1} Z_{i+1} \right)$, (b) $H_{4} = -\sum_i \left( Z_{i-1} X_i Z_{i+1} + Y_i Y_{i+1} \right)$, and (c) $H_{1} = -\sum_i \left( Z_{i-1} X_i Z_{i+1} + Z_i Z_{i+1} \right)$. 
  In (a), $L$ denotes the number of unit cells, corresponding to $2L$ spin sites, whereas in (b,c), $L$ denotes the number of spin sites. 
  All observables are evaluated at separation $r = L/4$ and centered in the chain, with $r$ measured in unit cells in (a) and lattice sites in (b,c); 
  their definitions are given in the End Matter. 
  Dotted lines indicate the reference power laws $\propto L^{-2\Delta}$. 
  (a) At the KT self-dual point of $H_{2}$, the odd/even-sublattice spin correlations $C_{\text{o/e},2}(L/4)$ and SPT string order parameter $O^Z_{\text{o/e},2}(L/4)$ decay as $L^{-1/4}$, while the PM string order parameter $O^I_{\text{o/e},2}(L/4)$ decay as $L^{-1/2}$. 
  The inset shows the energy splitting $\delta E \equiv E_2-E_1$ induced by symmetry-preserving transverse fields of strength $0.1$ at both ends, with a fit $\delta E(L) \propto L^{-1.005}$, consistent with the $1/L$ scaling of the bulk finite-size gap. 
  This $1/L$ splitting was also reported in Ref.~\cite{XiaoWang2023arxiv};
  it can be understood analytically after applying an open-chain SPT entangler.
  (b) For $H_{4}$, $C_Y(L/4)$, $O_{\text{SPT}}(L/4)$, and $O_{\text{PM}}(L/4)$ all decay as $L^{-3/4}$; see SM Sec.~\ref{sm3} for the explanation.
  The inset shows the lowest-lying eight energy levels $E_k$ for $L = 30$, with a unique ground state; see SM Sec.~\ref{sm3} for more discussions.
  (c) For $H_{1}$, $C_Z(L/4)$ and $O_{\text{SPT}}(L/4)$ decay as $L^{-1/4}$, whereas $O_{\text{PM}}(L/4)$ decays as $L^{-9/4}$.
  The inset shows the lowest-lying eight energy levels for $L = 32$, with a twofold ground-state degeneracy.}
  \label{fig:fig2}
\end{figure*}

\emph{\color{red} KT self-duality and symmetry-enriched criticality.}---Recent studies~\cite{YU20261, Verresen2021PRX} have uncovered a new class of QCPs, known as symmetry-enriched critical points (see the End Matter for a brief introduction), which are topologically distinct from conventional critical points in the sense that they cannot be 
smoothly connected without encountering a multicritical point or breaking the protecting symmetries.
As a prototypical example, we consider the $(1+1)$-dimensional $\mathbb Z_2 \times \mathbb{Z}_{2}$ cluster-Ising model, $H_{2}=-\sum_{i}\left(J_{2}'Z_{i-1}Z_{i+1}+g_{2}Z_{i-1}X_{i}Z_{i+1}\right)$, which is protected by independent $\mathbb Z_{2}^{\mathrm{even/odd}}$ spin-flip symmetries generated by $P_{\mathrm{even/odd}} = \prod_{i \in \mathrm{even/odd}} X_i$ on the even and odd sublattices.
Its phase diagram contains a continuous transition between an FM SSB and a cluster SPT phase [Fig.~\ref{fig:schematic}(a-2)], which are interchanged by the KT transformation, $Z_{i-1}X_iZ_{i+1} \leftrightarrow Z_{i-1}Z_{i+1}$.
The critical point $J_2' = g_2$ is therefore KT self-dual~\cite{KT1992PRB,Oshikawa_1992,Li2023PRB_b,Li2025SciPost,Parayil2024PRB}.
Under PBC, the KT transformation can be expressed as $U_\mathrm{KT} = \mathcal{D}^{\dagger}_{\mathrm{odd}}\mathcal{D}^{\dagger}_{\mathrm{even}} U_2 \mathcal{D}_{\mathrm{even}} \mathcal{D}_{\mathrm{odd}}$ with $U_\mathrm{KT}^2 = \frac{1+P_{\mathrm{even}}}{2} \frac{1+P_{\mathrm{odd}}}{2}$~\cite{Li2023PRB_b,Lu2024SciPost}, where $\mathcal{D}_{\mathrm{even/odd}}$ denotes gauging the $\mathbb{Z}_{2}^{\mathrm{even/odd}}$ symmetry on the even/odd sublattice, and $U_2=\prod_i \text{CZ}_{i,i+1}$ is the cluster SPT entangler, which implements stacking with a $\mathbb{Z}_2^{\mathrm{even}}\times\mathbb{Z}_2^{\mathrm{odd}}$ SPT phase.
Thus, the KT self-duality is noninvertible and consists of gauging both $\mathbb{Z}_2^{\mathrm{even/odd}}$ symmetries followed by SPT stacking~\cite{Lu2024SciPost}, giving rise to a symmetry-enriched $\mathrm{Ising}^{2}$ critical point that is topologically distinct from the critical point of the two-copy transverse-field Ising chain.
In fact, this distinction is encoded in the scaling properties of bulk nonlocal string operators carrying different symmetry charges, rather than by the existence of topological edge modes, which constitute only one possible consequence of the underlying topology.
As illustrated in Fig.~\ref{fig:fig2}(a), the charged SPT string operators on both even and odd sublattices exhibit the slowest algebraic decay, with scaling dimension $1/8$, while the charge-neutral PM string operators decay more rapidly with scaling dimension $1/4$, indicating that the critical point is topologically nontrivial~\cite{Verresen2021PRX}; see End Matter for the definitions of the observables.
Nevertheless, the corresponding edge modes are unstable against the symmetry-preserving boundary perturbation, since their energy splitting exhibits the same $\sim 1/L$ scaling as the bulk finite-size gap, as shown in the inset of Fig.~\ref{fig:fig2}(a).
By contrast, consider the critical Hamiltonian $H_4 = -\sum_i (Z_{i-1}X_{i}Z_{i+1}  + Y_iY_{i+1})$, which does not possess KT self-duality.
As shown in Fig.~\ref{fig:fig2}(b), the open-boundary energy spectrum exhibits no ground-state topological degeneracy.
More importantly, the FM connected correlation, as well as the SPT and PM string operators, all exhibit identical algebraic decay with exponents $3/4$, demonstrating that the critical point is topologically trivial. 
Further discussion of the subtle properties of $H_4$ is provided in SM Sec.~\ref{sm3}.

At first sight, these observations might suggest that the KT self-duality alone is sufficient to enforce symmetry-enriched quantum criticality. KT self-duality alone, however, is not sufficient, as demonstrated by the critical Hamiltonian $H_{3} = -\sum_{i} \left( Z_{i-1}Z_{i+1} + X_{i} + Z_{i-1}X_{i}Z_{i+1} \right)$, which possesses the same KT self-duality but exhibits no protected topological degeneracy. The essential difference is that the associated SPT entangler $U_2$ is already an exact symmetry of the lattice Hamiltonian, satisfying $U_2^{-1}H_3U_2=H_3$. Even after introducing irrelevant lattice perturbations that explicitly break the $U_2$ symmetry so that $H_3\neq U_2^{-1}H_3U_2$ microscopically, the two Hamiltonians still flow to the same infrared CFT and carry identical charges in all twisted sectors. We therefore conclude that KT self-duality alone may not be sufficient to guarantee symmetry-enriched quantum criticality. The missing ingredient is that the associated SPT entangler must not be promoted to an emergent symmetry of the infrared CFT, thereby endowing the critical theory with distinct twisted-sector charges. 

We also remark that antiunitary symmetries can also enrich quantum criticality. A representative example is the $(1+1)$-dimensional cluster-Ising model protected by $\mathbb{Z}_{2}\times\mathbb{Z}_{2}^{T}$ symmetry~\cite{Verresen2021PRX}, $H_{1}=-\sum_{i}\left(J_{2}Z_{i}Z_{i+1}+g_{2}Z_{i-1}X_{i}Z_{i+1}\right)$, where $\mathbb{Z}_{2}$ denotes the global spin-flip symmetry and $\mathbb{Z}_{2}^{T}$ the spinless time-reversal symmetry. Although $H_{1}$ does not possess the KT self-duality defined above, it can be mapped onto the transverse-field Ising model through the SPT entangler $U_2$, whose critical point is KW self-dual. The resulting critical point can therefore be viewed as the conventional Ising critical point dressed by the SPT entangler, thereby playing a role analogous to the KT self-duality and realizing symmetry-enriched Ising criticality~\cite{Verresen2021PRX,Yu2022PRL}. As shown in Fig.~\ref{fig:fig2}(c), this is manifested by the SPT string operator, which exhibits the slowest algebraic decay with scaling dimension $1/8$, together with a robust twofold ground-state degeneracy~\cite{Verresen2021PRX} in the open-boundary energy spectrum. The corresponding field-theoretic description, however, remains an interesting open problem. Therefore, throughout this work, we restrict our discussion to the $\mathbb{Z}_n^{\mathrm{even}} \times \mathbb{Z}_n^{\mathrm{odd}}$ qudit model.

\emph{\color{red} Generalization to  $\mathbb{Z}_n \times \mathbb{Z}_n$ cluster-qudit chains.}---Beyond the exactly solvable $\mathbb{Z}_2\times\mathbb{Z}_2$ cluster-Ising chains, the same mechanism extends naturally to the $\mathbb{Z}_n\times\mathbb{Z}_n$ cluster-qudit chains, which host $\mathbb{Z}_n \times \mathbb{Z}_n$ SSB and SPT phases. Concretely, we consider the cluster-qudit Hamiltonian, $H_{\mathbb{Z}_n\text{-cluster}} = -\sum_{i}(J_n R^{\dagger}_{2i-1}R_{2i+1} + J_n R^{\dagger}_{2i}R_{2i+2} + g_{n}R^{\dagger}_{2i-2}M_{2i-1}R_{2i}+g_nR_{2i-1}M_{2i}R^{\dagger}_{2i+1}+\mathrm{h.c.})$, protected by $\mathbb{Z}_n^{\mathrm{even}}\times\mathbb{Z}_n^{\mathrm{odd}}$ symmetry. 
This model admits a natural $\mathbb{Z}_n$ generalization of the KT transformation, $U_{\mathrm{KT},n}$, which exchanges the cluster and symmetry-breaking interactions, 
\begin{equation}
\label{eq2}
\begin{split}
&U_{\mathrm{KT},n}R_{2i-1}M_{2i}R_{2i+1}^{\dagger} = 
R_{2i}R_{2i+2}^{\dagger}U_{\mathrm{KT},n},\\
&U_{\mathrm{KT},n}R^{\dagger}_{2i}M_{2i+1}R_{2i+2} = 
R_{2i+1}R^{\dagger}_{2i+3}U_{\mathrm{KT},n},
\end{split}
\end{equation}
thereby rendering the critical point $J_n=g_n$ KT self-dual, which admits a natural field-theoretic interpretation as a combination of gauging the $\mathbb{Z}_n^{\mathrm{even}}\times\mathbb{Z}_n^{\mathrm{odd}}$ symmetry and stacking with a $\mathbb{Z}_n^{\mathrm{even}}\times\mathbb{Z}_n^{\mathrm{odd}}$ SPT. Under PBC, the resulting KT transformation is promoted to a noninvertible symmetry~\cite{Parayil2024PRB} at criticality and obeys the same criterion established for the $\mathbb{Z}_2$ case: symmetry-enriched quantum criticality emerges at the KT self-dual point only when the $\mathbb{Z}_n$ SPT entangler~\cite{Parayil2024PRB} does not become a symmetry of the infrared CFT, thereby assigning distinct charges to different twisted sectors. The infrared fixed point is therefore described by the symmetry-enriched $\mathbb{Z}_n^{\mathrm{even}} \times \mathbb{Z}_n^{\mathrm{odd}}$ CFT. Consequently, for $n\le4$, both lattice realizations exhibit a direct symmetry-enriched QCP between the $\mathbb{Z}_n\times\mathbb{Z}_n$ SPT and SSB phases [Fig.~\ref{fig:schematic}(a-2)]. For $n\ge5$, the same KT self-duality instead stabilizes an extended topological Luttinger liquid, namely a gapless SPT phase with central charge $c=2$, separating the two gapped phases [Fig.~\ref{fig:schematic}(b-2)].

\begin{figure}[htbp]
  \centering
  \includegraphics[width=1.0\linewidth]{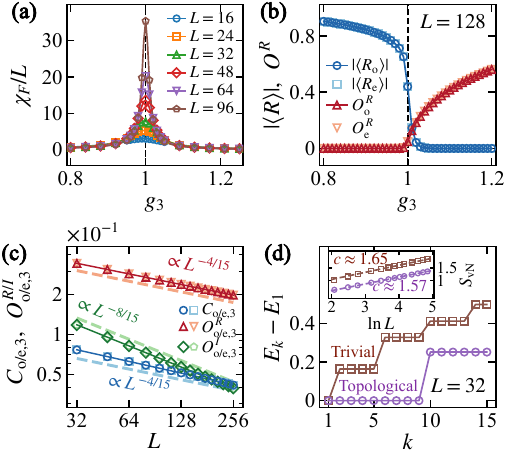}
  \caption{Numerical results for the $\mathbb{Z}_{3} \times \mathbb{Z}_{3}$ cluster-qutrit chain under open boundary conditions with fixed $J_3 = 1$. 
  Here $L$ denotes the number of unit cells, corresponding to $2L$ qutrit sites, and $\mathrm{o/e}$ labels the odd/even sublattice. 
  (a) Fidelity susceptibility per unit cell, $\chi_F/L$, versus $g_3$ at $h_3 = 0$ for various system sizes.
  The peak sharpens near the KT self-dual point $g_3 = 1$, marked by the vertical dashed line.
  (b) Local order parameters $|\langle R_{\text{o/e},\text{center}}\rangle|$ and SPT string order parameters $O^R_{\text{o/e},\text{full}}$ versus $g_3$ at $h_3 = 0$ for $L = 128$.
  The label ``center'' denotes the central unit cell, while ``full'' denotes a string whose endpoints extend to the two boundaries of the chain.
  The local and string order parameters characterize the SSB and SPT phases, respectively.
  (c) Finite-size scaling of the connected spin correlations $C_{\text{o/e},3}(r)$, SPT string order parameters $O^R_{\text{o/e},3}(r)$, and PM string order parameters $O^I_{\text{o/e},3}(r)$ at the critical point $(h_3,g_3) = (0,1)$.
  These observables are evaluated at separation $r = L/4$, measured in unit cells, and centered in the chain.
  Dashed lines indicate $L^{-4/15}$ decay for $C_{\text{o/e},3}(L/4)$ and $O^R_{\text{o/e},3}(L/4)$, and $L^{-8/15}$ decay for $O^I_{\text{o/e},3}(L/4)$, corresponding to scaling dimensions $\Delta = 2/15$ and $4/15$, respectively.
  (d) Lowest-lying fifteen energy levels $E_k - E_1$ for $L = 32$ at the topological $(h_3,g_3) = (0,1)$ and trivial $(1,0)$ critical points.
  The topological case exhibits a ninefold ground-state degeneracy, whereas the trivial case has a unique ground state.
  The inset shows the half-chain von Neumann entanglement entropy $S_{\text{vN}}(L)$ versus $\ln L$.
  Fits to $S_{\text{vN}}(L) = (c/6) \ln L + \text{const.}$ yield $c \approx 1.57$ and $1.65$ for the topological and trivial cases, respectively, consistent with the total central charge $c = 2 \times 4/5$ of two three-state Potts sectors. Definitions of all physical quantities are provided in the End Matter.}
  \label{fig:fig3}
\end{figure}

We now test these predictions numerically using the $\mathbb{Z}_3^{\mathrm{even}} \times \mathbb{Z}_3^{\mathrm{odd}}$ cluster-qutrit chain as a representative example (see SM Sec.~\ref{sm4} for analogous results for $n=4$ and $5$). As shown in Fig.~\ref{fig:fig3}(a), the peak of the fidelity susceptibility converges to the self-dual point $g_3/J_3=1$ with increasing system size, confirming that the transition is pinned by the KT self-duality. The continuous nature of the transition is further demonstrated in Fig.~\ref{fig:fig3}(b): the local order parameter $R_{\mathrm{even/odd}}$ and the SPT string order parameter $O_{\mathrm{even/odd}}^{R}$ (see the End Matter for the explicit definitions) both vanish continuously upon approaching the critical point from the SSB and SPT phases, respectively. At criticality, both the SPT string operator $O^{R}_{\mathrm{even/odd}}$ and the PM string operator $O^{I}_{\mathrm{even/odd}}$, carrying different $\mathbb{Z}_3^{\mathrm{odd/even}}$ symmetry charges, exhibit algebraic decay with distinct scaling dimensions. The SPT string operator has scaling dimension $2/15$, identical to that of the $\mathbb{Z}_3$ spin operator $R_{\mathrm{even/odd}}$, as required by the KT self-duality, whereas the PM string operator has scaling dimension $4/15$.
Remarkably, the latter is exactly twice the former, in precise agreement with the $\mathbb{Z}_3$ $\mathrm{Potts}^2$ CFT prediction for the nonlocal primary fields $\mu$ ($\mu'$) and their composite $\mu\mu'$~\cite{prembabu2025multicriticalitypurelygaplessspt}; also see SM Sec.~\ref{sm2}.
Moreover, the present KT self-dual point is not invariant under the SPT entangler and is therefore expected to be topologically distinct from the conventional KW self-dual point. As confirmed in Fig.~\ref{fig:fig3}(d): although both critical points have the same central charge, $c\simeq1.6$, only the KT self-dual point exhibits a stable ninefold ground-state degeneracy, whereas the conventional KW self-dual point does not, identifying the former as a symmetry-enriched $\mathbb{Z}_3$ $\mathrm{Potts}^2$ CFT.

\emph{\color{red} Higher dimensions.}---Finally, we briefly discuss how the proposed KT self-duality naturally extends beyond one dimension. Higher-dimensional systems host a much broader landscape of symmetry-breaking and SPT phases than their one-dimensional counterparts. As a natural extension of the present framework, we can consider continuous transitions between subsystem $\mathbb{Z}_n\times\mathbb{Z}_n$ SPT phases~\cite{You2018PRB,Devakul2018PRB,Cao2023SciPost} and spontaneous subsystem symmetry-breaking phases on two sublattices, whose construction closely parallels the one-dimensional case~\cite{Parayil2024PRB,maity2025webnoninvertibledualities21}. The corresponding KT transformation maps the subsystem $\mathbb{Z}_n\times\mathbb{Z}_n$ SPT phase onto two copies of the hypercubic $\mathbb{Z}_n$ clock model. From the field-theoretic perspective, this transformation remains noninvertible and can be understood as a higher-dimensional analogue of the KW transformation~\cite{Parayil2024PRB,Su2026SciPost}
followed by SPT stacking. A complementary extension concerns transitions from trivial or SSB phases to mixed SPT phases protected jointly by an ordinary $\mathbb{Z}_n^{(0)}$ symmetry and a $\mathbb{Z}_n^{(1)}$ one-form symmetry. For $n=2$, a generalized KT transformation has been explicitly constructed in a $(2+1)$-dimensional lattice gauge theory, establishing a connection between SSB and mixed SPT phases~\cite{Choi2025SciPost}. A crucial difference from one dimension is that the SPT stacking operation in higher dimensions can involve a broader range of SPT resources, including ordinary, subsystem, and higher-form SPT phases~\cite{Yoshida2016PRB,verresen2024higgscondensatessymmetryprotectedtopological,Jian2021PRB}. Consequently, the associated KT transformations also become considerably richer, substantially enlarging the landscape of possible KT self-dual points and symmetry-enriched criticality.

\emph{\color{red} Concluding remarks.}--- \color{black} To summarize, we have established KT self-duality as the criterion for identifying a broad class of one-dimensional symmetry-enriched QCPs separating SPT and SSB phases. In close analogy with the role of KW self-duality in conventional criticality, the noninvertible KT self-duality not only pins the critical point but also endows it with nontrivial topological features whenever the associated SPT entangler is not promoted to an emergent symmetry of the infrared CFT. We substantiate this mechanism numerically in $\mathbb{Z}_n \times \mathbb{Z}_{n}$ cluster-qudit chains and further argue that it naturally extends beyond one dimension, thus providing a unified framework for exploring symmetry-enriched criticality across dimensions.

An intriguing open question is whether alternative mechanisms exist that can realize symmetry-enriched quantum criticality. Developing scalable numerical methods for higher-dimensional realizations remains an outstanding challenge. Recent theoretical proposals and numerical studies of $\mathbb{Z}_{n}$ qudits in Rydberg atom arrays~\cite{Benhemou2023PRR,verresen2023quantumisingmodel,yeo2026entanglementstructuremathbbznphase}, together with the ability to engineer cluster-type interactions on the same platform~\cite{samajdar2026threebodyinteractionsrydberglattices, Petiziol2021PRL}, provide a promising route toward exploring these phenomena.






\textit{\color{red} Acknowledgement}:
We thank Saranesh Prembabu, Yunqin Zheng, Weicheng Ye and Junchen Rong for helpful discussions.
X.-J. Yu was supported by the National Natural Science Foundation of China (Grant No.~12405034) and a start-up grant from Eastern Institute of Technology, Ningbo. The work of S.Y. is supported by China Postdoctoral Science Foundation (Certificate Number: 2024M752760).

\bibliographystyle{apsrev4-2}
\let\oldaddcontentsline\addcontentsline
\renewcommand{\addcontentsline}[3]{}
\bibliography{main.bib}

\appendix
\section{\large{End Matter}}
\twocolumngrid

\textit{Brief introduction to symmetry-enriched quantum criticality.}---Quantum critical points can be further enriched by global symmetries, giving rise to symmetry-enriched quantum criticalities (SEQCs)~\cite{Verresen2021PRX,YU20261}.
Specifically, symmetry enrichment can split a single universality class into several distinct topological sectors.
SEQCs sharing the same infrared conformal field theory (CFT) therefore cannot be adiabatically connected to one another without either passing through a multicritical point or breaking the protecting symmetry.

To date, most known examples of SEQCs were identified empirically at continuous phase transitions between symmetry-protected topological (SPT) phases and spontaneous symmetry-breaking (SSB) phases.
A paradigmatic example is provided by the quantum phase transitions in the transverse-field Ising and cluster-Ising chains, described by $H_\text{TFI}$ (with $J_{2} = h_{2}$) and $H_{1}$ (with $J_{2} = g_{2}$), respectively, as defined in the main text.
Both quantum critical points are described by the $(1+1)$-dimensional Ising CFT.
The order and disorder fields of the Ising CFT, $\sigma$ and $\mu$,
have equal scaling dimensions $\Delta_{\sigma} = \Delta_{\mu} = 1/8$~\cite{Verresen2021PRX,Yu2022PRL}.
Here, $\sigma$ is the magnetic order parameter, which can be represented by the microscopic lattice operator $\sigma_{i} \propto Z_{i}$, whereas $\mu$ is the Kramers--Wannier dual string order parameter of the symmetric phase.
Explicitly, for the transverse-field Ising chain, $\mu_{i} \propto \prod_{m = -\infty}^{i-1} X_{m}$, while for the cluster-Ising chain, $\tilde{\mu}_{i} \propto \left[ \prod_{m = -\infty}^{i-2} X_{m} \right] Y_{i-1} Z_{i}$.
These two nonlocal string operators can be interpreted as symmetry-flux operators carrying distinct time-reversal charges, satisfying $T \mu_{i} T = +\mu_{i}$ and $T \tilde{\mu}_{i} T = -\tilde{\mu}_{i}$, respectively~\cite{Verresen2021PRX}.
The two-point functions $O_{\text{PM}}(r) \equiv \langle \mu_{i} \mu_{i+r} \rangle$ and $O_{\text{SPT}}(r) \equiv \langle \tilde{\mu}_{i} \tilde{\mu}_{i+r} \rangle$ decay with the same exponent $1/4$ and exhibit the slowest algebraic decay at the respective quantum critical points.
Consequently, the two quantum critical points are topologically inequivalent, with the time-reversal charge of the nonlocal disorder operator serving as the topological invariant~\cite{Verresen2021PRX}.
In particular, the cluster-Ising critical point exhibits a twofold ground-state degeneracy protected by the $\mathbb{Z}_2 \times \mathbb{Z}_2^T$ protecting symmetry: under symmetry-preserving boundary perturbations, the finite-size energy splitting of the topological degeneracy decays as $\sim L^{-14}$~\cite{Verresen2021PRX}, parametrically faster than the bulk finite-size gap (e.g., $\sim L^{-1}$) of the Ising CFT, rendering the topological degeneracy stable.

\textit{Observables for the spin-$1/2$ lattice models.}---Here we provide the definitions of the connected spin-spin correlation functions, paramagnetic (PM) string order parameters, and SPT string order parameters shown in Fig.~\ref{fig:fig2} of the main text.

For $H_{2}$, the relevant observables can be defined separately with respect to the odd and even sublattices.
The connected spin-spin correlation functions on the two sublattices are defined as (here $\text{o/e}$ denotes $\text{odd/even}$)
\begin{equation}
  \begin{split}
  C_{\text{o},2}(|i-j|) & \equiv \langle Z_{2i-1} Z_{2j-1} \rangle - \langle Z_{2i-1} \rangle \langle Z_{2j-1} \rangle \, , \\
  C_{\text{e},2}(|i-j|) & \equiv \langle Z_{2i} Z_{2j} \rangle - \langle Z_{2i} \rangle \langle Z_{2j} \rangle \, .
  \end{split}
\end{equation}
The corresponding SPT string order parameters are
\begin{equation}
  \begin{split}
  O_{\text{o},2}^{Z}(|i-j|) & \equiv \left\langle Z_{2i} \left[ \prod_{m=i+1}^{j} X_{2m-1} \right] Z_{2j} \right\rangle \, , \\
  O_{\text{e},2}^{Z}(|i-j|) & \equiv \left\langle Z_{2i-1} \left[ \prod_{m=i}^{j-1} X_{2m} \right] Z_{2j-1} \right\rangle \, .
  \end{split}
\end{equation}
The associated PM string order parameters, denoted by $O_{\text{o},2}^{I}(r)$ and $O_{\text{e},2}^{I}(r)$, are obtained by replacing the endpoint $Z$ operators in the above expressions with identity operators.

For $H_{1}$ and $H_{4}$, the global $\mathbb{Z}_{2}$ spin-flip symmetry acts on the entire chain, and there is therefore no distinction between odd- and even-sublattice observables.
The connected spin-spin correlation functions are defined as
\begin{equation}
  \begin{split}
  C_{Z}(|i-j|) & \equiv \langle Z_{i} Z_{j} \rangle - \langle Z_{i} \rangle \langle Z_{j} \rangle \, , \\
  C_{Y}(|i-j|) & \equiv \langle Y_{i} Y_{j} \rangle - \langle Y_{i} \rangle \langle Y_{j} \rangle \, .
  \end{split}
\end{equation}
The PM and SPT string order parameters are given by $O_{\text{PM}}(r)$ and $O_{\text{SPT}}(r)$, respectively, as defined in the preceding section, with the same definitions applying to both $H_{1}$ and $H_{4}$.

\textit{Lattice model for $\mathbb{Z}_{n} \times \mathbb{Z}_{n}$ qudit chains.}---In this work, we consider a $\mathbb{Z}_{n} \times \mathbb{Z}_{n}$ qudit chain with $L$ unit cells, whose Hamiltonian interpolates among the PM, SSB, and SPT phases,
\begin{equation}
  \begin{split}
  \label{eq:ZnZn_model}
  H_{\mathbb{Z}_{n}} & = h_{n} H_{\text{PM}}^{(n)} + J_{n} H_{\text{SSB}}^{(n)} + g_{n} H_{\text{SPT}}^{(n)} \, , \\
  H_{\text{PM}}^{(n)} & = -\sum_{i} \left( M^{}_{2i-1} + M^{}_{2i} + \text{h.c.} \right) \, , \\
  H_{\text{SSB}}^{(n)} & = -\sum_{i} \left( R^{\dagger}_{2i-1} R^{}_{2i+1} + R^{\dagger}_{2i} R^{}_{2i+2} + \text{h.c.} \right) \, , \\
  H_{\text{SPT}}^{(n)} & = -\sum_{i} \left( R^{\dagger}_{2i-2} M^{}_{2i-1} R^{}_{2i} + R^{}_{2i-1} M^{}_{2i} R^{\dagger}_{2i+1} + \text{h.c.} \right) \, .
  \end{split}
\end{equation}
Each unit cell labeled by $i$ carries a finite local Hilbert space spanned by $\ket{\Gamma_{2i-1},\Gamma_{2i}} \equiv \ket{\Gamma_{2i-1}} \otimes \ket{\Gamma_{2i}}$, where $(\Gamma_{2i-1},\Gamma_{2i}) \in \mathbb{Z}_{n} \times \mathbb{Z}_{n}$ with $\Gamma_{j} = 0, 1, \dots, n-1$, and the total Hilbert space is the tensor product of local Hilbert spaces defined on all unit cells.
The shift and clock operators are defined as $M_{j} \ket{\Gamma_j} = \ket{\Gamma_j+1}$ mod $n$ and $R_j \ket{\Gamma_j} = \eta_{n}^{\Gamma_j} \ket{\Gamma_j}$, where $\eta_{n} \equiv \exp\left( 2\pi \text{i}/n \right)$.
The model respects two independent $\mathbb{Z}_n$ global symmetries generated by $P_\text{o} = \prod_{i=1}^{L} M^{}_{2i-1}$ and $P_\text{e} = \prod_{i=1}^{L} M^{}_{2i}$, which act on the odd and even sublattices, respectively.

At $g_{n} = 0$ and $h_{n} = J_{n}$, the system realizes a topologically trivial quantum critical point described by
the product of the critical theories of two $\mathbb{Z}_{n}$ clock chains,
whereas at $h_{n} = 0$ and $g_{n} = J_{n}$, it realizes a symmetry-enriched quantum criticality, which is characterized by the same theory.
In the main text, we focus only on the Ising ($n = 2$) and three-state Potts ($n = 3$) cases as illustrative examples.
For $n = 3$, the shift and clock operators read
\begin{equation}
  M = \begin{pmatrix}
    0 & 0 & 1 \\
    1 & 0 & 0 \\
    0 & 1 & 0
  \end{pmatrix} \, , \qquad R = \begin{pmatrix}
    1 & 0 & 0 \\
    0 & \text{e}^{2\pi\text{i}/3} & 0 \\
    0 & 0 & \text{e}^{4\pi\text{i}/3}
  \end{pmatrix} \, ,
\end{equation}
so that $M |0\rangle = |1\rangle$, $M |1\rangle = |2\rangle$, and $M |2\rangle = |0\rangle$.
Here the group elements of $\mathbb{Z}_{3}$ are labeled by $0$, $1$, $2$, with $0$ denoting the identity element.

To probe the distinct quantum phases of the $\mathbb{Z}_n \times \mathbb{Z}_n$ qudit chains, we compute local spin order parameters and nonlocal string order parameters, each of which can be defined separately on the odd and even sublattices.
The FM order is characterized by the connected spin-spin correlation functions,
\begin{align}
  C_{\text{o},n}(|i-j|) & \equiv \langle R^{\dagger}_{2i-1} R^{}_{2j-1} \rangle - \langle R^{\dagger}_{2i-1} \rangle \langle R^{}_{2j-1} \rangle \, , \\
  C_{\text{e},n}(|i-j|) & \equiv \langle R^{\dagger}_{2i} R^{}_{2j} \rangle - \langle R^{\dagger}_{2i} \rangle \langle R^{}_{2j} \rangle\, .
\end{align}
The corresponding SPT string order parameters are defined as
\begin{align}
  O_{\text{o},n}^{R}(|i-j|) & \equiv \left\langle R^{\dagger}_{2i} \left[\prod_{m=i+1}^{j} M^{}_{2m-1} \right] R^{}_{2j} \right\rangle \, , \\
  O_{\text{e},n}^{R}(|i-j|) & \equiv \left\langle R^{}_{2i-1} \left[ \prod_{m=i}^{j-1} M^{}_{2m} \right] R^{\dagger}_{2j-1} \right\rangle \, .
\end{align}
The PM string order parameters are obtained by replacing the endpoint operators $R^{\dagger}_{}$ and $R^{}_{}$ with the identity operator, and are denoted by $O_{\text{o},n}^{I}(r)$ and $O_{\text{e},n}^{I}(r)$.

\textit{Numerical methods.}---To accurately determine the quantum critical point of the generalized $\mathbb{Z}_{n} \times \mathbb{Z}_{n}$ qudit models, we employ the ground-state fidelity susceptibility~\cite{Gu2010IJMPB,Gu2008PRB}.
For example, taking $J_{n} = 1$ as the energy unit and $h_{n} = 0$, we can tune the dimensionless coupling ratio $\lambda \equiv g_{n}/J_{n}$.
The ground-state fidelity between two nearby values of the control parameter, $\lambda$ and $\lambda + \delta\lambda$, is defined as $F(\lambda, \lambda + \delta\lambda) \equiv \left| \langle \psi_0(\lambda) | \psi_0(\lambda+\delta\lambda) \rangle \right|$, where $|\psi_0(\lambda)\rangle$ denotes the ground state at $\lambda$.
The fidelity susceptibility characterizes the leading quadratic response of the fidelity to an infinitesimal parameter variation,
\begin{equation}
  \label{eq:fidelity_susceptibility}
  \chi_{F}(\lambda) \equiv \lim_{\delta\lambda \rightarrow 0} \frac{2\left[1-F(\lambda,\lambda+\delta\lambda)\right]}{(\delta\lambda)^2} \, .
\end{equation}
In numerical practice, we set $\delta \lambda = 10^{-3}$.
Near a continuous quantum critical point, $\chi_{F}$ exhibits singular finite-size scaling, manifested as a pronounced size-dependent peak, whereas its density $\chi_{F}/L$ remains regular away from the criticality.
We therefore identify the finite-size pseudocritical point $\lambda_{c}(L)$ from the peak position of $\chi_{F}/L$, which converges to the thermodynamic-limit critical point $\lambda_{c}$ as the system size $L$ increases.

To further characterize the critical behavior, we calculate the bipartite von Neumann entanglement entropy.
Partitioning the system into two complementary subsystems $\mathcal{L}$ and $\mathcal{R}$, the reduced density matrix of subsystem $\mathcal{L}$, containing $l$ unit cells, is obtained by tracing out $\mathcal{R}$, $\rho_{\mathcal{L}} \equiv \text{Tr}_{\mathcal{R}} \! \left( \ket{\psi_{0}} \! \bra{\psi_{0}} \right)$.
The corresponding entanglement entropy is defined as $S_{l} \equiv -\text{Tr} \! \left( \rho_{\mathcal{L}} \ln\rho_{\mathcal{L}} \right)$, which quantifies the bipartite entanglement between $\mathcal{L}$ and $\mathcal{R}$.
For a one-dimensional critical system described by a CFT, the ground-state entanglement entropy exhibits universal logarithmic scaling, with a prefactor determined by the central charge $c$.
For a finite-size system with periodic boundary conditions, the Calabrese--Cardy formula reads~\cite{Pasquale_Calabrese_2004,di1997conformal,ginsparg1988appliedconformalfieldtheory}, $S_l \sim \frac{c}{3} \ln\left[ \frac{L}{\pi} \sin \left( \frac{\pi l}{L} \right) \right] + S^{\prime}$, where $S^{\prime}$ is a nonuniversal constant.
For the open boundary conditions employed throughout this work, the half-chain entanglement entropy at criticality scales as
\begin{equation}
  \label{eq:EE_OBC}
  S_\text{vN}(L) \equiv S_{L/2} \sim \frac{c}{6} \ln L + S^{\prime\prime} \, ,
\end{equation}
where $S^{\prime\prime}$ is another nonuniversal constant.
We therefore extract the central charge $c$ from the slope of a linear fit of $S_\text{vN}(L)$ versus $\ln L$.
Throughout this work, ``$\ln$'' denotes the natural logarithm.

All numerical results presented in this work are obtained using finite-size density-matrix renormalization group (DMRG) calculations within the matrix product state (MPS) framework~\cite{White1992PRL,SCHOLLWOCK201196,Vidal2003PRL}, implemented with the ITENSOR \verb|C++| library~\cite{itensor}.
We retain MPS bond dimensions up to $\chi_{\text{max}}=1024$.
The ground-state optimization is deemed converged when the relative change in energy between two consecutive sweeps falls below $10^{-8}$.
All simulations are performed with open boundary conditions.

From the optimized ground states, we evaluate the fidelity susceptibility, half-chain entanglement entropy, correlation functions, and string order parameters defined above.
Low-energy excited states are obtained iteratively by performing variational MPS optimizations in subspaces orthogonal to the previously determined lower-energy states, implemented by adding suitable projector penalties to the Hamiltonian~\cite{Stoudenmire2012ARCMP}.

Some of the models considered here possess a symmetry-enriched critical point with a degenerate ground-state manifold under open boundary conditions.
To select a definite state within this manifold, we apply a weak boundary field.
For example, for the $\mathbb{Z}_{n} \times \mathbb{Z}_{n}$ cluster-qudit chain, we add $-h_\text{b} \left( R_{1}^{} + R_{1}^{\dagger} + R_{2L}^{} + R_{2L}^{\dagger} \right)$, which acts only on the two end sites.
The boundary field is applied when computing ground-state quantities, including the connected correlation functions, string order parameters, half-chain entanglement entropy, and fidelity susceptibility.
It is not included in the calculations of the low-lying excitation spectrum.

\let\addcontentsline\oldaddcontentsline
\onecolumngrid

\clearpage
\newpage

\widetext

\begin{center}
\textbf{\large Supplemental Material for ``Symmetry-Enriched Criticality from Kennedy–Tasaki Self-Duality''}
\end{center}

\maketitle

\counterwithout{equation}{section}
\renewcommand{\thefigure}{S\arabic{figure}}
\setcounter{figure}{0}
\renewcommand{\theequation}{S\arabic{equation}}
\setcounter{equation}{0}
\renewcommand{\thesection}{\Roman{section}}
\setcounter{section}{0}
\setcounter{secnumdepth}{4}

\addtocontents{toc}{\protect\setcounter{tocdepth}{0}}
{
\tableofcontents
}

\section{Explicit expression for KW and KT transformations}
\label{sm1}
Following Ref.~\cite{Parayil2024PRB}, the generalized Kramers--Wannier (KW) transformation for the $\mathbb{Z}_n$ quantum clock model with an $n$-dimensional qudit on each lattice site can be written under periodic boundary conditions (PBC) as
\begin{align}
\mathbf{D}_{(n)}
&\equiv
\mathbf{P}_{(n)}
\tilde{\mathbf{D}}_{(n)}
\mathbf{P}_{(n)}
\nonumber\\
&=
\mathbf{P}_{(n)}
\prod_{j=1}^{L-1}
\left[
\left(
\sum_{k=0}^{n-1}
c(k)M_j^k
\right)
\left(
\sum_{k=0}^{n-1}
\tilde{c}(k)
\left(R_jR_{j+1}^{\dagger}\right)^k
\right)
\right]
\left(
\sum_{k=0}^{n-1}
c(k)M_L^k
\right)
\mathbf{P}_{(n)},
\label{eq:D_N0}
\end{align}
where
\begin{align}
Q_{n} \equiv \prod_{j=1}^{L} M_{j}^{} \, , \quad \, \mathbf{P}_{(n)} = \frac{1}{n} \sum_{k=0}^{n-1} Q_{n}^{k} 
\end{align}
is the projector onto the $\mathbb{Z}_n$-symmetric sector. The coefficients appearing in Eq.~\eqref{eq:D_N0} are
\begin{align}
\tilde{c}(k)
=
\frac{\eta_{n}^{-\frac{k(n+k)}{2}}}{\sqrt{n}},
\qquad
c(k)
=
\frac{\eta_{n}^{\frac{k(n-k)}{2}}}{\sqrt{n}}.
\end{align}

The generalized KW transformation satisfies the following algebraic relations:
\begin{subequations}
\begin{align}
\mathbf{D}_{(n)}M_j
&=
R_jR_{j+1}^{\dagger}\mathbf{D}_{(n)},
\\
\mathbf{D}_{(n)}M_j^{\dagger}
&=
R_j^{\dagger}R_{j+1}\mathbf{D}_{(n)},
\\
\mathbf{D}_{(n)}R_jR_{j+1}^{\dagger}
&=
M_{j+1}\mathbf{D}_{(n)},
\\
\mathbf{D}_{(n)}R_j^{\dagger}R_{j+1}
&=
M_{j+1}^{\dagger}\mathbf{D}_{(n)}.
\end{align}
\label{eq:D_N}
\end{subequations}

At the self-dual point $\lambda=1$, $\mathbf{D}_{(n)}$ commutes with the Hamiltonian. Moreover, it satisfies
\begin{align}
\mathbf{D}_{(n)}^{\dagger}\mathbf{D}_{(n)}
=
\mathbf{D}_{(n)}
\mathbf{D}_{(n)}^{\dagger}
=
\mathbf{P}_{(n)},
\end{align}
and
\begin{align}
\mathbf{D}_{(n)}^{2}
=
e^{2i\phi(n,L)}
\mathbf{T}_{\mathrm{clock}}
\mathbf{P}_{(n)}.
\end{align}
Here, $\mathbf{T}_{\mathrm{clock}}$ denotes the one-site lattice translation operator, and $e^{i\phi(n,L)}$ is an overall phase factor. Explicit values of $e^{i\phi(n,L)}$ for several small values of $n$ are listed in Table~I of Ref.~\cite{Parayil2024PRB}. 

Based on the generalized KW transformation introduced above, we now construct explicit expression of the KT transformation for the $\mathbb{Z}_n\times\mathbb{Z}_n$ cluster-qudit chain. We first define the $\mathbb{Z}_n$ cluster SPT entangler by introducing the qudit generalization of the controlled-$Z$ gate,
\begin{align}
\mathrm{CZ}^{k}_{c,t}
\equiv
\sum_{m=0}^{n-1}
|m\rangle_c\langle m|
\otimes
Z_t^{km},
\qquad
\mathrm{CZ}^{\dagger}_{c,t}
\equiv
\mathrm{CZ}^{\,n-1}_{c,t},
\end{align}
where $c$ and $t$ denote the control and target qudits, respectively. The corresponding cluster SPT entangler is given by
\begin{align}
\mathcal{T}
=
\prod_{i=1}^{L}
\mathrm{CZ}_{2i,2i-1}
\mathrm{CZ}^{\dagger}_{2i,2i+1},
\qquad
\mathcal{T}^{n}=1,
\end{align}
which maps the $\mathbb{Z}_n\times\mathbb{Z}_n$ SPT Hamiltonian to its trivial symmetric counterpart.

The generalized KT transformation is then defined as
\begin{align}
U_{\mathrm{KT},n}
=
\mathcal{T}^{\dagger}
\mathbf{D}_{n,\mathrm{odd}}
\mathbf{D}_{n,\mathrm{even}}
\mathcal{T}^{\dagger},
\end{align}
where $\mathbf{D}_{n,\mathrm{odd}}$ and $\mathbf{D}_{n,\mathrm{even}}$ denote the generalized KW transformations acting on the odd and even sublattices, respectively. By construction, $U_{\mathrm{KT},n}$ maps the $\mathbb{Z}_n\times\mathbb{Z}_n$ SPT phase onto two decoupled $\mathbb{Z}_n$ SSB phases.

For the $d$-dimensional hypercubic $\mathbb{Z}_n$ clock model, the generalized KW transformation can be constructed as
\begin{align}
\label{eqs12}
\mathbf{D}_{n}^{(d)}
\equiv
\mathbf{P}_{n}^{(d)}
\tilde{\mathbf{D}}_{n,d}
\mathbf{H}_{n}^{\otimes(d)}
\tilde{\mathbf{D}}_{n,d-1}
\mathbf{H}_{n}^{\otimes(d)}
\cdots
\tilde{\mathbf{D}}_{n,1}
\mathbf{P}_{n}^{(d)},
\end{align}
where $\tilde{\mathbf{D}}_{n,\mu}$ ($\mu=1,\cdots,d$) denotes the generalized string operator acting along the $\mu$th lattice direction. For $n=2$, it reduces to
\begin{align}
\tilde{\mathbf{D}}
=
\left(
\prod_{j=1}^{L-1}
e^{i\frac{\pi}{4}X_j}
e^{i\frac{\pi}{4}Z_jZ_{j+1}}
\right)
e^{i\frac{\pi}{4}X_L},
\end{align}
defined on a one-dimensional line. Here,
$\mathbf{H}_{n}^{\otimes(d)}$ denotes the tensor product of the generalized Hadamard operator $\mathbf{H}_n$ over all sites of the $d$-dimensional lattice, while $\mathbf{P}_{n}^{(d)}$ projects onto the sector preserving the subsystem $\mathbb{Z}_n$ symmetries along every straight lattice line.

At the self-dual point, $\mathbf{D}_{n}^{(d)}$ commutes with the Hamiltonian. Up to an overall normalization, it furthermore satisfies
\begin{align}
\left(\mathbf{D}_{n}^{(d)}\right)^2
\propto
\mathbf{C}\,
\mathbf{T}_{(1,\ldots,1)}
\mathbf{P}_{n}^{(d)},
\end{align}
where $\mathbf{T}_{(1,\ldots,1)}$ denotes the diagonal lattice translation by one unit along all spatial directions, and $\mathbf{C}$ is the charge-conjugation operator.

The corresponding $d$-dimensional KT transformation is then defined as
\begin{align}
U_{\mathrm{KT},n}^{(d)}
=
\mathbf{D}_{1,n}^{(d)\dagger}
\mathbf{D}_{2,n}^{(d)\dagger}
\mathcal{T}^{(d)}
\mathbf{D}_{2,n}^{(d)}
\mathbf{D}_{1,n}^{(d)},
\end{align}
where $\mathbf{D}_{1,n}^{(d)}$ and $\mathbf{D}_{2,n}^{(d)}$ are the generalized KW transformations acting on the two sublattices, respectively, and $\mathcal{T}^{(d)}$ denotes the $d$-dimensional $\mathbb{Z}_n$ cluster SPT entangler, which can incorporate multiple types of SPT stacking.

\section{Factor-of-two difference between PM and SPT string decay exponents in $\mathbb{Z}_n\times\mathbb{Z}_n$ models}
\label{sm2}

In this section, we explain the factor-of-two difference between the decay exponents of the PM and SPT string order parameters in the $\mathbb{Z}_n \times \mathbb{Z}_n$ qudit models studied here.
At $(J_n, g_n, h_n) = (1, 0, 1)$, the Hamiltonian in Eq.~\eqref{eq:ZnZn_model} consists of two completely decoupled KW self-dual chains on the odd and even sublattices.
The KT self-dual critical chain at $(J_n, g_n, h_n) = (1, 1, 0)$ is related to this trivial model by the SPT entangler.
The two models therefore share the same two-copy CFT, while their string operators have different scaling dimensions.

For the trivial critical point, the SPT string differs from the PM string only by two spin operators on the opposite sublattice.
Since the two sublattices are decoupled, the odd-sublattice SPT string operator factorizes exactly as
\begin{equation}
  O_{\text{o},n}^{R}(|i-j|) = \left\langle \prod_{m=i+1}^{j} M^{}_{2m-1} \right\rangle_{\text{o}} \left\langle R_{2i}^{\dagger} R_{2j}^{} \right\rangle_{\text{e}} = O_{\text{o},n}^{I}(r) \times \left\langle R_{2i}^{\dagger} R_{2j}^{} \right\rangle_{\text{e}} \, .
  \label{eq:string_factorization}
\end{equation}
The even-sublattice string factorizes in the same way, with the roles of the two sublattices exchanged.
Therefore, we have
\begin{equation}
  O_{\text{o/e},n}^{I}(r) \propto r^{-2\Delta_n} \, , \quad O_{\text{o/e},n}^{R}(r) \propto r^{-4\Delta_n} \, .
  \label{eq:string_scaling_trivial}
\end{equation}
Here $\Delta_{n} = 1/8$, $2/15$, $1/8$, and $1/10$, respectively, for $n = 2$, $3$, $4$, and $5$~\cite{XiongJie2017PRB}.

The SPT entangler defined in Sec.~\ref{sm1} leaves the spin operators invariant and dresses the PM string with nontrivial endpoint operators.
Thus, at the symmetry-enriched critical points, the SPT string order parameter becomes the two-point function of the disorder field in a single sector.
The PM string instead maps to a disorder-field two-point function multiplied by a spin correlator in the other sector.
The two string scaling dimensions are therefore interchanged
\begin{equation}
  O_{\text{o/e},n}^{R}(r) \propto r^{-2\Delta_n} \, , \quad O_{\text{o/e},n}^{I}(r) \propto r^{-4\Delta_n} \, .
  \label{eq:string_scaling_topological}
\end{equation}

For $n = 5$, the above argument applies to the self-dual point within the gapless phase, where order- and disorder-field scaling dimensions coincide.
For the bulk observables at separation $r = L/4$, centered in the chain, these scaling dimensions give finite-size decays $L^{-2\Delta_n}$ and $L^{-4\Delta_n}$, consistent with Figs.~\ref{fig:fig2}(a), \ref{fig:fig3}(c), \ref{fig:sm_trivial_z3}, \ref{fig:sm_z4}(c,d), and \ref{fig:sm_z5}(c,d).

\section{More discussions on the model $H_4$}
\label{sm3}

\begin{figure}[t]
    \centering
    \includegraphics[width=0.7\textwidth]{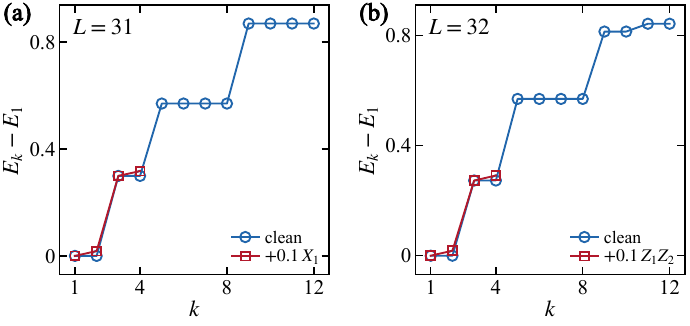}
    \caption{Low-lying energy spectra $E_k - E_1$ of $H_{4} = -\sum_i \left( Z_{i-1} X_i Z_{i+1} + Y_i Y_{i+1} \right)$ under open boundary conditions for (a) $L = 31$ and (b) $L = 32$, where $L$ is the number of spin sites.
    Blue circles show the lowest-lying twelve levels of the unperturbed Hamiltonian, with a twofold ground-state degeneracy for both sizes.
    Red squares show the lowest-lying four levels after adding symmetry-preserving boundary perturbations $H_{\text{bdy}} = \lambda X_1$ in (a) and $H_{\text{bdy}} = \lambda Z_1 Z_2$ in (b), with $\lambda = 0.1$\,.
    Both perturbations preserve the global spin-flip symmetry generated by $\prod_i X_i$ and lift the twofold ground-state degeneracy.
    The finite-size scaling of the induced splitting is shown in Fig.~\ref{fig:zxz_yy_splitting}.}
    \label{fig:zxz_yy_p3}
\end{figure}

In this section, we analyze the physics of the 1D lattice model governed by the Hamiltonian:
\begin{equation}
  H_{4} = - \sum_{i} Z_{i-1} X_{i} Z_{i+1} - \sum_{i} Y_{i} Y_{i+1} \, .
\label{eq:ZXZ_YY}
\end{equation}
Our primary aim is to elucidate the origin of its size-dependent ground-state degeneracy under open boundary conditions; 
see Fig.~\ref{fig:fig2}(b) in the main text and Fig.~\ref{fig:zxz_yy_p3} for concrete examples.
More importantly, we establish a clear distinction between this size-dependent degeneracy and the topologically protected degenerate edge modes typically found in gapless SPT states.
Under the SPT conjugation, the model transforms to $H_{4}' = -\sum_{i} X_{i} - \sum_{i} Z_{i} X_{i+1} X_{i+2} Z_{i+3}$, which does not possess the conventional KW duality.

To this end, we employ the Jordan--Wigner transformation defined as:
\begin{equation}
  \gamma_{2i-1} = \Bigg[ \prod_{j < i} X_{j} \Bigg] Z_{i} \, , \quad \text{and} \quad \gamma_{2i} = \Bigg[ \prod_{j < i} X_{j} \Bigg] Y_{i} \, .
\end{equation}
Under this mapping, the three-body and two-body interacting terms in Eq.~\eqref{eq:ZXZ_YY} transform as:
\begin{equation}
  Z_{i-1} X_{i} Z_{i+1} = \text{i} \gamma_{2i-2} \gamma_{2i+1} \quad \text{and} \quad Y_{i} Y_{i+1} = - \text{i} \gamma_{2i-1} \gamma_{2i+2} \, .
\end{equation}
Substituting these expressions back into the Hamiltonian yields
\begin{equation}
  H_{4}^\text{OBC} \to \tilde{H}_{4}^\text{OBC} = - \text{i} \sum_{k=1}^{2L-3} (-1)^{k} \gamma_{k} \gamma_{k+3} \, ,
\end{equation}
where $L$ denotes the total number of physical spins, corresponding to $2L$ Majorana modes.

Crucially, the hopping amplitudes in this Majorana representation occur strictly between modes separated by a distance of $3$. 
Consequently, the system completely decouples into three independent, noninteracting Majorana subchains: (1) $\{\gamma_1, \gamma_4, \gamma_7, \dots\}$, (2) $\{\gamma_2, \gamma_5, \gamma_8, \dots\}$, and (3) $\{\gamma_3, \gamma_6, \gamma_9, \dots\}$. 
Each subchain describes a gapless, free Majorana fermion chain, implying that the total system realizes a gapless state described by a $c = 3/2$ CFT.

\textbf{Size-dependent degeneracy under open boundary conditions.}
Under open boundary conditions, the degeneracy of the ground-state manifold is intimately tied to the chain length $L$ due to the completely decoupled nature of the three subchains. 
Since the total number of Majorana modes is partitioned among three subchains, we can classify the ground-state degeneracy based on the commensurability condition $L \bmod 3$: 
\begin{itemize}
  \item[(i)] For $L = 3l$: The total number of Majorana modes is an exact multiple of $6$. As a result, all three subchains contain an even number of modes, leading to a unique ground state.
  \item[(ii)] For $L = 3l+1$: The distribution of the $6l+2$ Majorana modes leaves exactly two subchains (specifically, subchains $1$ and $2$) with an odd number of modes.
  \item[(iii)] For $L = 3l+2$: Similarly, the distribution results in exactly two subchains possessing an odd number of Majorana modes, which are subchains $2$ and $3$. 
\end{itemize}
In the latter two cases ($L \neq 3l$), each of the two odd-length subchains hosts an isolated zero-energy mode. 
Together, these two modes form a single complex fermion, yielding an exact two-fold ground-state degeneracy; see Fig.~\ref{fig:zxz_yy_p3}.

\begin{figure}[t]
    \centering
    \includegraphics[width=0.7\textwidth]{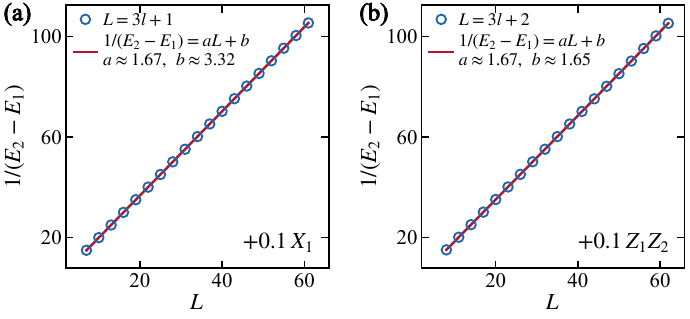}
    \caption{Finite-size scaling of the ground-state energy splitting $\delta E \equiv E_2 - E_1$ induced by symmetry-preserving boundary perturbations in $H_{4}$ under open boundary conditions: (a) for $L = 3l + 1$ with $H_{\text{bdy}} = \lambda X_1$ and (b) for $L = 3l+2$ with $H_{\text{bdy}} = \lambda Z_1 Z_2$.
    Here $L$ denotes the number of spin sites, $l$ is an integer, and $\lambda = 0.1$ in both panels.
    Blue circles show the inverse splitting $1/\delta E(L)$, and red solid lines are linear fits to $1/\delta E(L) = a L + b$, yielding $(a,b) \approx (1.67, 3.32)$ in (a) and $(1.67, 1.65)$ in (b).
    Thus, the induced splitting scales asymptotically as $\delta E(L) \propto L^{-1}$ in both cases, on the same scale as the bulk finite-size gap.
    }
    \label{fig:zxz_yy_splitting}
\end{figure}

\textbf{Energy splitting of the degeneracy under symmetry-preserving boundary perturbations.}
It is crucial to distinguish this two-fold degeneracy from the robust topological edge modes that characterize genuine gapless SPT phases. 
Here, we investigate the finite-size energy splitting of the observed two-fold degeneracy by introducing boundary perturbations that preserve the global fermion parity symmetry, which corresponds to the global spin-flip symmetry generated by $\prod_{i} X_{i}$.

For $L = 3l+1$, the two zero modes contributing to the two-fold degeneracy originate from subchains $1$ and $2$. 
We can introduce a boundary perturbation of the form $H_\text{bdy}^{(3l+1)} = \lambda X_1 \sim \text{i} \lambda \gamma_1 \gamma_2$, which directly couples subchains $1$ and $2$ at the left boundary. 
Conversely, for $L = 3l+2$, the two zero modes reside on subchains $2$ and $3$. 
Accordingly, we can introduce a different boundary perturbation, $H_\text{bdy}^{(3l+2)} = \lambda Z_{1} Z_{2} \sim \text{i} \lambda \gamma_2 \gamma_3$, which couples subchains $2$ and $3$ at the left boundary.

As shown in Fig.~\ref{fig:zxz_yy_p3}, these two boundary perturbations lift the twofold ground-state degeneracy for $L = 31$ and $L = 32$, respectively. 
Furthermore, we studied the finite-size scaling of the energy splitting as shown in Fig.~\ref{fig:zxz_yy_splitting}. 
In both cases, the energy splitting induced by the boundary perturbation decays as $\delta E \equiv E_{2} - E_{1} \sim 1/(aL+b)$. 
This power-law decay of the splitting is asymptotically identical to the scaling of the bulk energy-gap closing, which scales as $\delta E_\text{bulk} \sim 1/L$.

In conclusion, the emergence of the two-fold degeneracy in the model~\eqref{eq:ZXZ_YY} is highly sensitive to the microscopic lattice details (i.e., $L \bmod 3$) and can be immediately lifted by boundary perturbations that even strictly preserve the global $\mathbb{Z}_{2}$ spin-flip symmetry. 
This explicitly confirms that the system is not a gapless SPT state, as its boundary modes lack robust symmetry protection.

\textbf{Common decay exponent $3/4$ of the bulk observables.}
The three-subchain decomposition also explains the common decay exponent $3/4$ of $C_Y(r)$, $O_{\text{PM}}(r)$, and $O_{\text{SPT}}(r)$ shown in Fig.~\ref{fig:fig2}(b) in the main text.
In a ground state preserving the global spin-flip symmetry, $\langle Y_i \rangle = 0$, so that $C_Y(r) = \langle Y_i Y_{i+r} \rangle$.
Using the string operators defined in the End Matter and the Jordan--Wigner transformation above, we obtain
\begin{equation}
    Y_i Y_{i+r} = \prod_{k=i}^{i+r-1} \left( -\text{i} \gamma_{2k-1} \gamma_{2k+2} \right) \, , \quad 
    \mu_i \mu_{i+r} \propto \prod_{k=i}^{i+r-1} \left( \text{i} \gamma_{2k-1} \gamma_{2k} \right) \, , \quad
    \tilde{\mu}_i \tilde{\mu}_{i+r} \propto \prod_{k=i}^{i+r-1} \left( \text{i} \gamma_{2k-2} \gamma_{2k+1} \right) \, .
  \label{eq:H4_correlator_majoranas}
\end{equation}
For $r = 3m$, as in the data shown in Fig.~\ref{fig:fig2}(b) in the main text, regrouping the Majoranas by their indices modulo $3$ turns each expression into a product of three strings, each containing $2m$ consecutive Majoranas in one subchain contributing a factor $\propto r^{-1/4}$.
Therefore, the total magnitudes obey
\begin{equation}
  C_Y(r) ,\ O_{\text{PM}}(r) ,\ O_{\text{SPT}}(r) \propto \left( r^{-1/4} \right)^3 = r^{-3/4} \, .
\end{equation}
For the bulk observables evaluated at separation $r = L/4$ and centered in the chain, the corresponding finite-size scaling is $L^{-3/4}$, as observed in Fig.~\ref{fig:fig2}(b) in the main text.

\section{Additional numerical results for $\mathbb{Z}_n \times \mathbb{Z}_{n}$ qudit chains}
\label{sm4}

\begin{figure}[t]
    \centering
    \includegraphics[width=0.4\textwidth]{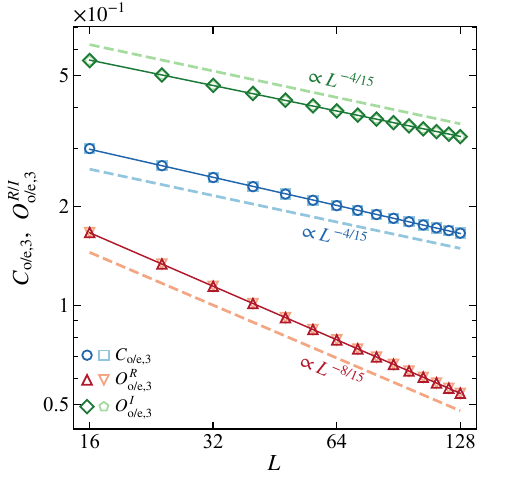}
    \caption{Finite-size scaling at the topologically trivial $\mathbb{Z}_{3}^{\text{even}} \times \mathbb{Z}_{3}^{\text{odd}}$ critical point of Eq.~\eqref{eq:ZnZn_model}, with $J_3 = h_3 = 1$ and $g_3 = 0$, under open boundary conditions. 
    The connected spin correlation functions $C_{\text{o/e},3}(r)$, PM string order parameters $O^{I}_{\text{o/e},3}(r)$, and SPT string order parameters $O^{R}_{\text{o/e},3}(r)$ are evaluated at separation $r = L/4$ and centered in the chain, where $L$ is the number of unit cells and $\text{o/e}$ denotes the odd/even sublattice. 
    Dashed lines indicate the expected power-law decays: $C_{\text{o/e},3}(L/4)$ and $O^{I}_{\text{o/e},3}(L/4)$ scale as $L^{-4/15}$, whereas $O^{R}_{\text{o/e},3}(L/4)$ scales as $L^{-8/15}$. 
    Thus, the PM string operators have the smaller scaling dimension $\Delta = 2/15$, while the SPT string operators have $\Delta = 4/15$. 
    At the topologically nontrivial critical point in Fig.~\ref{fig:fig3}(c) in the main text, these scaling dimensions are interchanged.}
    \label{fig:sm_trivial_z3}
\end{figure}

\begin{figure}[t]
    \centering
    \includegraphics[width=0.6\textwidth]{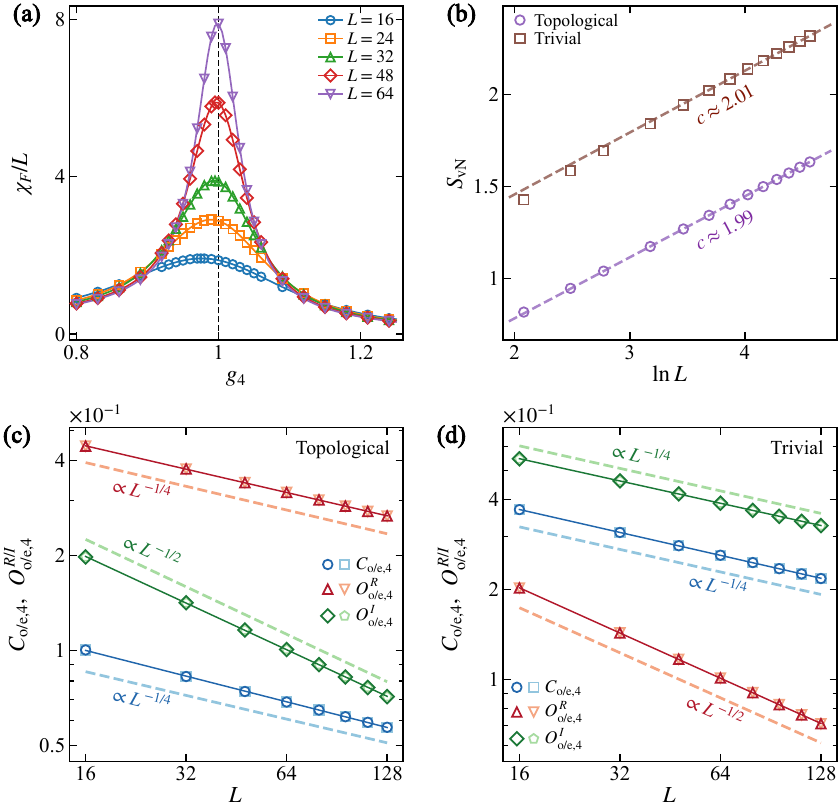}
    \caption{Numerical results for the $\mathbb{Z}_{4}^{\text{even}} \times \mathbb{Z}_{4}^{\text{odd}}$ qudit chain in Eq.~\eqref{eq:ZnZn_model} under open boundary conditions with $J_4 = 1$. 
    The topologically nontrivial and trivial critical points correspond to $(h_4,g_4) = (0,1)$ and $(1,0)$, respectively. 
    (a) Fidelity susceptibility per unit cell, $\chi_F/L$, as a function of $g_4$ at $h_4 = 0$ for various system sizes. 
    The peak sharpens near the KT self-dual point $g_4 = 1$, marked by the vertical dashed line. 
    (b) Half-chain von Neumann entanglement entropy $S_{\text{vN}}(L)$ versus $\ln L$ at the two critical points. 
    Dashed linear fits to $S_{\text{vN}}(L) = (c/6)\ln L + \text{const.}$ yield $c \approx 1.99$ and $2.01$ for the topologically nontrivial and trivial cases, respectively.
    (c,d) Finite-size scaling of the connected spin correlations $C_{\text{o/e},4}(r)$, SPT string order parameters $O^{R}_{\text{o/e},4}(r)$, and PM string order parameters $O^{I}_{\text{o/e},4}(r)$ at the topologically nontrivial and trivial critical points, respectively.
    Here $L$ is the number of unit cells, and all observables are evaluated at separation $r = L/4$ and centered in the chain.
    Dashed lines indicate $L^{-1/4}$ and $L^{-1/2}$ decays, corresponding to scaling dimensions $\Delta = 1/8$ and $1/4$.
    The spin correlations decay as $L^{-1/4}$ in both cases, while the SPT and PM string operators exchange their scaling dimensions, distinguishing the two critical points despite their common central charge.}
    \label{fig:sm_z4}
\end{figure}

\begin{figure}[t]
    \centering
    \includegraphics[width=0.6\textwidth]{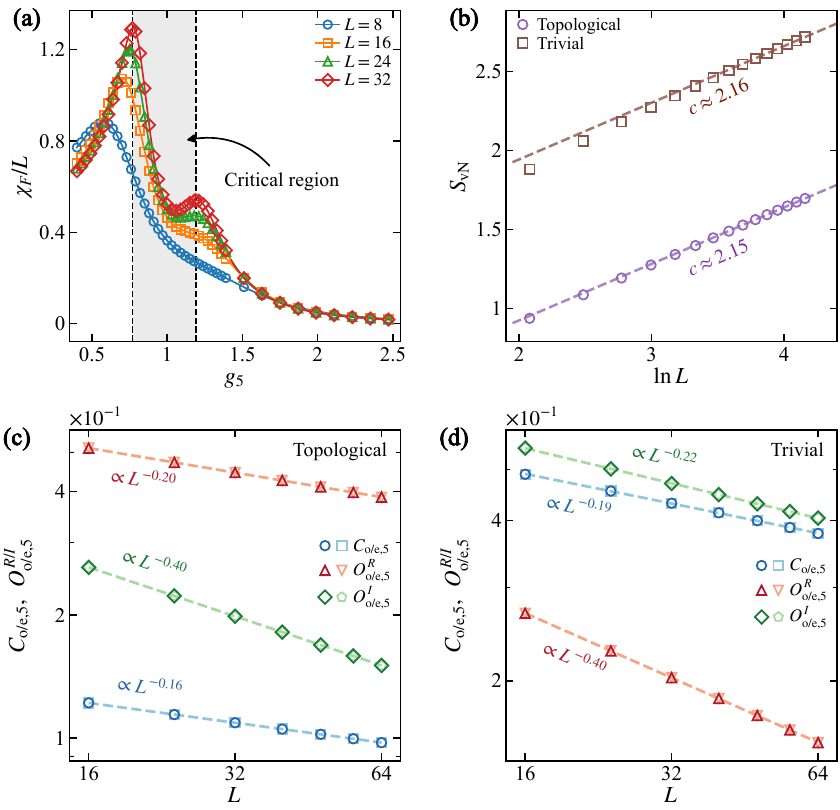}
    \caption{Numerical results for the $\mathbb{Z}_{5}^{\text{even}} \times \mathbb{Z}_{5}^{\text{odd}}$ qudit chain in Eq.~\eqref{eq:ZnZn_model} under open boundary conditions with $J_5 = 1$.
    (a) Fidelity susceptibility per unit cell, $\chi_F/L$, as a function of $g_5$ at $h_5 = 0$ for various system sizes.
    The two-peak structure is consistent with an intermediate extended gapless (floating) phase between the SSB and SPT phases.
    (b) Half-chain von Neumann entanglement entropy $S_{\text{vN}}(L)$ versus $\ln L$ at the KT self-dual point $(h_5,g_5) = (0,1)$ and its KW self-dual counterpart $(1,0)$.
    Dashed linear fits to $S_{\mathrm{vN}}(L) = (c/6)\ln L + \text{const.}$ give $c \simeq 2$ in both cases.
    (c,d) Finite-size scaling of the connected spin correlations $C_{\text{o/e},5}(r)$, SPT string order parameters $O^{R}_{\text{o/e},5}(r)$, and PM string order parameters $O^{I}_{\text{o/e},5}(r)$ in the topologically nontrivial and trivial cases, respectively, at the same self-dual parameters as in (b).
    Here $L$ is the number of unit cells, and all observables are evaluated at separation $r = L/4$ and centered in the chain.
    Dashed power-law fits give decay exponents $(0.16, 0.20, 0.40)$ for $(C, O^R, O^I)$ in (c) and $(0.19, 0.40, 0.22)$ in (d).
    These finite-size estimates show that the SPT string order parameters decay more slowly in the topologically nontrivial case, while the PM string order parameters decay more slowly in the trivial case.
    The theoretically predicted scaling dimensions should be $\Delta = 1/10$ and $1/5$~\cite{XiongJie2017PRB,Li2015PRB}, corresponding to decay exponents $2\Delta = 1/5$ and $2/5$;
    deviations of the fitted exponents from these theoretical values reflect finite-size effects within the floating phase.}
    \label{fig:sm_z5}
\end{figure}

In this section, we supplement the results shown in the main text by comparing physical observables of the topologically trivial and nontrivial $\mathbb{Z}_n \times \mathbb{Z}_n$ qudit chains for $n=3, 4$, and $5$.
We use the Hamiltonian in Eq.~\eqref{eq:ZnZn_model}, set $J_n=1$, and compare the KW self-dual parameters $(h_n,g_n) = (1,0)$ with the KT self-dual parameters $(0,1)$.
All calculations are performed under open boundary conditions.
Here $L$ denotes the number of unit cells, and the connected spin correlations and string order parameters are evaluated at separation $r = L/4$, centered in the chain.
For an operator with scaling dimension $\Delta$, a finite-size decay proportional to $L^{-2\Delta}$ will be observed.
Our comparison focuses on the scaling dimensions of the PM and SPT string operators, which distinguish the two critical points even when their central charges coincide.

\textbf{Topologically trivial criticality for $n = 3$.}---At $(h_3,g_3) = (1,0)$, the Hamiltonian reduces to two decoupled critical three-state Potts chains on the odd and even sublattices.
As shown in Fig.~\ref{fig:sm_trivial_z3}, the connected spin correlations $C_{\text{o/e},3}(L/4)$ and PM string order parameters $O^I_{\text{o/e},3}(L/4)$ both decay as $L^{-4/15}$, corresponding to $\Delta = 2/15$.
The SPT string order parameters $O^R_{\text{o/e},3}(L/4)$ instead decay as $L^{-8/15}$, with $\Delta = 4/15$.
At the KT self-dual critical point studied in the main text, the PM and SPT string operators exchange their scaling dimensions [see Fig.~\ref{fig:fig3}(c) in the main text], while the spin operator retains $\Delta = 2/15$.
Thus, the slowest-decaying string operator is charge neutral in the trivial critical point and carries a nontrivial symmetry charge in the topologically nontrivial critical point.
This comparison provides a characteristic signature of their topological distinction, complementary to the open-boundary energy spectra in Fig.~\ref{fig:fig3}(d) in the main text.
For chains comprising an integer number of complete unit cells, as in Fig.~\ref{fig:fig3}(d) of the main text, the ground state is ninefold degenerate.
Adding an even-sublattice site at the left boundary gives a half-integer number of unit cells and reduces the degeneracy to threefold.
The same boundary change reduces the degeneracy from $n^2$ to $n$ for $n = 2$, $4$, and $5$.

\textbf{Symmetry-enriched criticality for $n = 4$.}---For $n = 4$, along the SSB-SPT interpolation with $h_4 = 0$ and $J_4 = 1$, the fidelity susceptibility develops a single peak that sharpens near $g_4 = 1$ as the system size increases [Fig.~\ref{fig:sm_z4}(a)], consistent with a direct transition at the KT self-dual point.
The half-chain entanglement entropy follows the logarithmic scaling, giving $c \approx 1.99$ for the topological critical point and $c \approx 2.01$ for its trivial counterpart [Fig.~\ref{fig:sm_z4}(b)].
These values are consistent with a total central charge $c = 2$, accounting for both sublattice sectors.

Despite this common central charge, the two critical points differ in their string operators.
At $(h_4,g_4) = (0,1)$, $C_{\text{o/e},4}(L/4)$ and $O^R_{\text{o/e},4}(L/4)$ decay as $L^{-1/4}$, whereas $O^I_{\text{o/e},4}(L/4)$ decays as $L^{-1/2}$ [Fig.~\ref{fig:sm_z4}(c)].
At $(h_4,g_4) = (1,0)$, $C_{\text{o/e},4}(L/4)$ and $O^I_{\text{o/e},4}(L/4)$ exhibit the slower $L^{-1/4}$ decay, while $O^R_{\text{o/e},4}(L/4)$ exhibits the faster $L^{-1/2}$ decay [Fig.~\ref{fig:sm_z4}(d)].
The PM and SPT string operators therefore exchange the scaling dimensions $1/8$ and $1/4$, extending the distinction observed for $n = 3$ to $n = 4$.

\textbf{Extended symmetry-enriched gapless phase for $n = 5$.}---For $n = 5$, the fidelity susceptibility exhibits two peaks separated by an intermediate region [Fig.~\ref{fig:sm_z5}(a)], consistent with the extended gapless phase illustrated in Fig.~\ref{fig:schematic}(b-2) in the main text.
In this case, the KT self-dual point lies within the gapless phase, rather than identifying an isolated transition between the SSB and SPT phases.
At the self-dual parameters, the half-chain entanglement entropy fits yield $c \approx 2$ for both the topologically trivial and nontrivial critical points [Fig.~\ref{fig:sm_z5}(b)], consistent with a total central charge $c = 2$, corresponding to two gapless $c = 1$ sectors.

Figures~\ref{fig:sm_z5}(c) and \ref{fig:sm_z5}(d) show that the distinction in string operators persists in this extended gapless regime.
At the KT self-dual point, power-law fits give decay exponents approximately $0.20$ for $O^R_{\text{o/e},5}(r)$ and $0.40$ for $O^I_{\text{o/e},5}(r)$;
at the KW self-dual point, however, the corresponding values are approximately $0.40$ and $0.22$\,.
The spin correlations decay algebraically in both cases, with fitted exponents approximately $0.16$ and $0.19$, respectively.
These estimates retain the exchange of the slowly and rapidly decaying string operators.
Together with the $n = 2$, $3$, and $4$ results, this comparison supports the distinction between KT and KW self-dual critical points through the symmetry charges of the dominant string operators, both at isolated critical points and within an extended gapless phase.

\end{document}